\documentclass{iopart}

\eqnobysec

\usepackage{iopams}

\usepackage{wrapfig}
\usepackage{graphicx}
\usepackage{hyperref}

\usepackage{iopams_add} 

\newcommand{\jmmid}[1]{\mathrm{#1}}
\newcommand{\jmmvec}[1]{\boldsymbol{#1}}
\newcommand{\vecgrad}{\jmmvec{\nabla}}
\newcommand{\scakB}{k_\jmmid{B}}
\newcommand{\jmmabs}[1]{{\lvert #1 \rvert}}

\newcommand{\ket}[1]{\mathinner{|{#1}\rangle}}

\newcommand{\bok}[3]{\mathinner{\langle #1 | #2 | #3 \rangle}}

\newcommand{\BraKet}[2]{\left< #1 \vert #2 \right>}
\newcommand{\BOK}[3]{\left< #1 \left| #2 \right| #3 \right>}

\newcommand{\CASTEP}{\href{http://www.castep.org}{\textsc{Castep}}}

\begin{document}

\title[Electron-hole pairs during the adsorption dynamics of O$_2$ on Pd(100)]
{Electron-hole pairs during the adsorption dynamics of O$_2$ on Pd(100) -- Exciting or not?}

\author{J\"org Meyer$^{1,2}$ and Karsten Reuter$^{1,2}$}

\address{$^1$ 
Theory Department, Fritz-Haber-Institut der Max-Planck-Gesellschaft, 
Faradayweg 4-6, 14195 Berlin, Germany}
\address{$^2$ 
Department Chemie, Technische Universit{\"a}t M{\"u}nchen, 
Lichtenbergstrasse 4, 85747 Garching, Germany}

\ead{joerg.meyer@ch.tum.de}

\begin{abstract}
During the exothermic adsorption of molecules at solid surfaces dissipation of the released energy occurs via the excitation of electronic and phononic degrees of freedom. For metallic substrates the role of the non-adiabatic electronic excitation channel has been controversially discussed, as the absence of a band gap could favour an easy coupling to a manifold of electron-hole pairs of arbitrarily low energies. We analyse this situation for the highly exothermic showcase system of molecular oxygen dissociating at Pd(100), using time-dependent perturbation theory applied to first-principles electronic-structure calculations. For a range of different trajectories of impinging O$_2$ molecules we compute largely varying electron-hole pair spectra, which underlines the necessity to consider the high-dimensionality of the surface dynamical process when assessing the total energy loss into this dissipation channel. Despite the high Pd density of states at the Fermi level, the concomitant non-adiabatic energy losses nevertheless never exceed about 5\,\% of the available chemisorption energy. While this supports an electronically adiabatic description of the predominant heat dissipation into the phononic system, we critically discuss the non-adiabatic excitations in the context of the O$_2$ spin transition during the dissociation process.
\end{abstract}


\submitto{\NJP}
\maketitle

\section{Introduction}
\label{sect:intro}

Energy conversion at interfaces is at the centre of the rapidly growing field of basic energy science. This concerns desired conversions like solar to chemical energy, but also unavoidable by-products like the dissipation of chemical energy into heat. An atomistic understanding of the involved elementary processes is in all cases only just emerging, but is likely to question established views and macro-scale concepts. With respect to the dissipation of heat freed during exothermic surface chemical reactions such a prevailing view is that of a rapid equilibration with the local heat bath provided by the solid surface. The view is nurtured by the rare event dynamics resulting from the typically sizeably activated nature of surface chemical processes: While the actual elementary processes themselves take place on a picosecond time scale, times between such rare events are orders of magnitude longer. The understanding is then that in these long inter-process time spans any released chemical energy is rapidly distributed over sufficiently many surface phononic degrees of freedom to warrant a description in terms of a mere heat bath with defined local temperature. At the atomic scale this equilibration with the surface heat bath leads to an efficient loss of memory of the adsorbates about their history on the solid surface between subsequent rare events. This motivates the description in terms of a Markovian state dynamics that is e.g. underlying all present-day microkinetic formulations in heterogeneous catalysis \cite{dumesic93,chorkendorff03,reuter11}. In turn, the freed reaction energy only enters the determination of the local heat bath temperature, commonly achieved through a continuum heat balancing equation \cite{deutschmann08,matera09,matera10}.

\begin{wrapfigure}{R}{0.5\textwidth}
\centering
\includegraphics{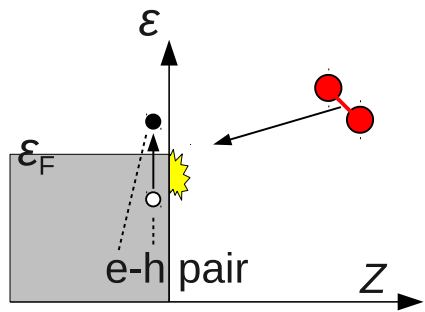}
\caption[Schematic illustration of electron-hole pair excitation.]
{Schematic illustration of electron-hole pair excitation during the impingement of a gas-phase molecule on a metal surface. An electron is excited from an occupied state below the Fermi level $\varepsilon_\jmmid{F}$ to an unoccupied state above, resulting in an excited state commonly referred to as electron-hole (e-h) pair.}
\label{fig1}
\end{wrapfigure}

In many cases this prevalent framework seems to allow for an accurate account e.g. of catalytic conversions. However, particularly for nanostructured surfaces and highly exothermic reactions it is presently unclear whether this effective description is sufficient. With respect to the exothermicity this suspicion comes from recalling that freed enthalpies can well be of the order of several eVs. This is e.g. generally the case for the dissociative adsorption of molecular oxygen at catalytically relevant transition metal surfaces, which is the specific surface chemical reaction we want to focus on here. Several eVs is an enormous amount of energy on the scale of phononic degrees of freedom, which calls for efficient dissipation channels to achieve the assumed quasi-instantaneous local equilibration. At metallic substrates electronic excitations could hereby potentially represent an important additional energy sink. Unfortunately, no consensus has hitherto been reached concerning the role of this additional channel for the gas-surface dynamics and subsequent energy dissipation. Among others Tully and coworkers have frequently stressed that the lack of a band gap in metallic systems should in principle allow for electronic excitations, namely electron-hole (e-h) pairs, of arbitrarily low energies to easily couple with the nuclear motion of a particle impinging from the gas-phase \cite{gordon92,gordon95,wodtke04,li05,cheng07a,cheng07b,shenvi09a,shenvi09b}. This is depicted schematically in \fref{fig1}. On the other hand, recent comparisons to experimental data for hydrogen and nitrogen molecules on metallic surfaces indicate that electronically adiabatic descriptions provided within the Born-Oppenheimer approximation (BOA) seem to describe the initial interaction dynamics governing the dissociation extremely well \cite{nieto06,juaristi08,luntz09}. As e.g. recently stressed by Juaristi \etal \cite{juaristi08} a proper high-dimensional description of the nuclear motion appears hereby significantly more important than the effect of e-h pair excitations.

For the particular case of oxygen dissociation the spin-triplet ground-state of gas-phase O$_2$ nevertheless brings in another aspect that could require to go beyond an adiabatic BOA treatment. With the spin of adsorbed O atoms at metal surfaces quenched, a spin transition needs to occur at some stage during adsorption process and could then well affect the gas-surface dynamics. The well-known paradigm system for this is O$_2$ at Al(111), where only an explicit account of a hindered triplet-singlet transition could reconcile first-principles dynamical simulations with the experimentally measured low sticking coefficient for thermal molecules \cite{behler05,behler08,carbogno08,carbogno10}. This non-adiabatic hindrance was traced back to the inefficiency of both coupling mechanisms generally discussed to relax the spin selection rules that suppress reactions like oxygen dissociation \cite{behler05,behler08,carbogno08,carbogno10,schwarz04}: The low mass number of Al leads to a small spin-orbit coupling and the low Al density of states (DOS) at the Fermi level prevents efficient spin quenching through the preferred tunnelling of minority spin electrons between substrate and adsorbate.

In this context the aim of the present study is to specifically address the electronic non-adiabaticity of the O$_2$ dissociation process at Pd(100). Apart from the obvious catalytic relevance, the specific motivation for the choice of the Pd substrate is its extraordinary high DOS at the Fermi level, which is among the highest known among the low-index transition metal surfaces 
\cite{arlinghaus81,elliott91,wu92,heinrichsmeier98}. 
Much in contrast to Al(111) this and the much higher mass number of Pd should favor an adiabatic spin transition, an expectation which we find confirmed by good agreement of a completely adiabatically computed initial sticking coefficient with existing experimental data. 
We are going to publish these results in detail elsewhere \cite{meyer11a}, thereby also addressing the most intriguing ``different dissociation properties'' of cartwheeling and helicoptering O$_2$ molecules on Pd(100).
These have been observed only very recently \cite{vattuone09,vattuone10}, most strikingly exemplifying the present limits of the aforementioned atomistic understanding of an elementary process like oxygen dissociation.
Here, however, the focus is on the excitation of e-h pairs instead, which according to the general arguments of Tully and coworkers \cite{gordon92,wodtke04,li05,cheng07a,cheng07b,shenvi09a,shenvi09b} should be particularly facilitated for such a high Fermi-level DOS system. 
The central objective is thus to assess how much of the total amount of $\sim 2.6$\,eV chemisorption energy ({\em vide infra}) in this system is dissipated into this electronic channel. In addition the computed spin-resolved e-h pair spectra will be critically analysed to look for features that relate to the adiabatic spin transition.

\section{Theory}
\label{sect:theory}

Several approaches of varying accuracy have been suggested to compute electronic excitation upon surface adsorption. At the high accuracy end is certainly the "direct" first-principles simulation of e-h pair spectra within time-dependent density-functional theory (DFT) and an Ehrenfest dynamics for the nuclei \cite{lindenblatt06a,lindenblatt06b,lindenblatt06c}. However, for the intended semi-quantitative estimate of electronic dissipation, and in view of the high computational demands imposed by the O$_2$ at Pd(100) system, more approximate treatments are appealing. Notwithstanding, the rather rough description of the substrate electronic structure in the time-dependent Newns-Anderson model developed by Mizielinski \etal \cite{mizielinski05,mizielinski07a,mizielinski07b,mizielinski08} raises some concerns. The same holds for approaches based on electronic friction theory \cite{gordon95,agliano75,persson82,hellsing84}, as their results can depend sensitively on the way how friction coefficients are calculated \cite{juaristi08,luntz09b,juaristi09,trail01,persson04,luntz05,luntz06}. More severely, already first applications within a forced oscillator model for the substrate electrons identify the proper description of spin transitions as an intrinsic shortcoming of electronic friction theory \cite{trail02,trail03}. For the present purposes we therefore opt for an approach introduced recently by Timmer and Kratzer (TK) \cite{timmer09}, which relies on perturbation theory applied to a time-dependent DFT framework. As the authors have demonstrated convincingly, it does not suffer from any troubles with the description of spin, and for the frequently studied reference system of hydrogen atoms at Al(111) very good agreement could be reached with the accurate time-dependent DFT/Ehrenfest benchmark data \cite{lindenblatt06c}. At the same time the approach is computationally very efficient, as it only requires less demanding ground-state DFT calculations as input.

In the following we recapitulate the TK approach. This is done self-contained and in considerable detail to highlight several aspects that deserve specific attention for the aspired application to the O$_2$ at Pd(100) system. Among those is the spin transition, which is why also a comparison to electronic friction theory is included. Furthermore, several improvements of the original TK approach are described. This concerns a numerically more efficient implementation to make the application to computationally more demanding systems like the present one tractable. As previous work has mainly been focused on purely one-dimensional model trajectories (of mono-atomic adsorbates), it also comprises the generalization to curved trajectories (of multi-atomic adsorbates). 

\subsection{Time-dependent perturbation theory}
\label{ssect:TDDFT}

The basis for the following derivation is a classical trajectory $\jmmvec{Q}(t)$ of a particle (a single atom or a molecule consisting of
$N$ atoms) from the gas phase impinging on the metal surface. In general, $\jmmvec{Q}$ is a $3N$ dimensional vector of Cartesian coordinates of the atoms, which might be replaced by a suitable reaction coordinate description. It is assumed that the particle collides with the surface and is reflected, which is commonly referred to as a ``full round trip''. Consequently, for very small and very large times $t$ the particle is assumed to be infinitely far away from the surface. This is meant in a physical sense such that the particle-surface interaction vanishes. In the philosophy of a perturbative approach, the effect of energy loss on the actual trajectory is neglected. Accordingly, a rigid surface that does not allow for any phonon excitation is considered, and energy dissipation into the electronic degrees of freedom of the surface is assumed to be sufficiently small not to significantly change the trajectory itself.

Thanks to the Runge-Gross theorem \cite{runge84}, the time dependent electronic many-body problem can be mapped onto an effective single particle Hamiltonian
\begin{equation}
\label{eq:ht}
  h^{\sigma}(t) = t_\jmmid{e} + v_\jmmid{TD,eff}^{\sigma}(t)
  \quad ,
\end{equation}
where $t_\jmmid{e}$ is the kinetic energy contribution and the time-dependent 
effective potential is given by
\begin{equation}
\label{eq:veff}
  v_\jmmid{TD,eff}^{\sigma}(t) = 
    v_\jmmid{TD,H}^{\sigma}(t) + 
    v_\jmmid{TD,xc}^{\sigma}(t) + 
    v_\jmmid{TD,ext}(t) 
  \; .
\end{equation}
$\sigma$ denotes collinear spin in two different channels within the framework of spin-DFT for collinear spin \cite{barth72}. $v_\jmmid{ext}^{\sigma}$ describes the electron-ion interaction and explicitly depends on time due to the motion of the impinging particle along its trajectory $\jmmvec{Q}(t)$. The dependence of the Hartree potential $v_\jmmid{H}^{\sigma}$ and exchange-correlation 
potential $v_\jmmid{xc}^{\sigma}$ on the time-dependent density $\rho(\jmmvec{r},t)$ is dropped in order to simplify readability.

At the ``beginning'' and the ``end'' of the full round trip, the (stationary) potential 
\begin{equation}
\label{eq:veff_inf}
  v_\jmmid{TD,eff}^{\sigma}(t) \;
    \xrightarrow[\lVert \jmmvec{Q}(t \rightarrow \pm \infty) \rVert_{2} \rightarrow \infty]
                {t \rightarrow \pm\infty} \;
  v_\jmmid{eff; \, non-int}^{\sigma}
\end{equation}
describes vanishing particle-surface interaction as mentioned before, yielding a (time-independent) Hamiltonian
\begin{equation}
\label{eq:h0}
  h^{\sigma}(t)
    \; \xrightarrow{t \rightarrow \pm\infty} \;
    t_\jmmid{e} + v_\jmmid{eff; \, non-int}^{\sigma}
  \quad \equiv \quad h_{(0)}^{\sigma}
  \quad .
\end{equation}

So far, no approximations have been made. To obtain a scheme that is more computationally tractable than the direct time-dependent DFT approach used e.g. by Lindenblatt and coworkers \cite{lindenblatt06a,lindenblatt06b,lindenblatt06c}, the time-dependent effective potential is now approximated by its counterparts in a ``series of snapshots'' of respective separate non-time-dependent ground
state problems along the considered trajectory $\jmmvec{Q}(t)$ as described by stationary DFT,
\begin{equation}
\label{eq:vQt}
\fl v_\jmmid{TD,eff}^{\sigma}(t) 
    \, \approx \,
      v_\jmmid{H}(\jmmvec{Q}(t)) + 
      v_\jmmid{xc}^{\sigma}(\jmmvec{Q}(t)) +  
      v_\jmmid{ext}(\jmmvec{Q}(t))
    \; \equiv \;
      \underbrace{v^{\sigma}(\jmmvec{Q}(t)) +  v_\jmmid{eff; \, non-int}^{\sigma}}
        _{\equiv v_\jmmid{eff}^{\sigma}(\jmmvec{Q}(t))}
  \quad ,
\end{equation}
with the interaction part of the approximated effective potential $v^{\sigma}(t)$ vanishing at the boundaries of the considered trajectory
\begin{equation}
\label{eq:v_inf}
  v^{\sigma}(t) \;
    \xrightarrow[\lVert \jmmvec{Q}(t \rightarrow \pm \infty) \rVert_{2} \rightarrow \infty]
                {t \rightarrow \pm\infty} \;
    0 \; .
\end{equation}
Following the arguments of TK \cite{timmer09}, the concomitant neglect of deviations from the instantaneous ground state charge density can only be justified for systems where the densities associated with excited electrons and holes are small compared to the overall 
charge density. From a theoretical point of view, this can, of course, be confirmed \emph{a posteriori} if the calculated electron-hole pair spectra are sufficiently small (\emph{vide infra}).

Equations~(\ref{eq:h0}) and (\ref{eq:vQt}) define an approximation for \eref{eq:ht},
\begin{equation}
\label{eq:hpert}
  h^{\sigma}(t) \approx h_{(0)} + v^{\sigma}(\jmmvec{Q}(t)) \; ,
\end{equation}
which is a quantum mechanical textbook example for a starting point of time-dependent perturbation theory: The unperturbed Hamiltonian $h_{(0)}$ gets acted on by a (implicitly) time- (and spin-) dependent perturbation potential $v^{\sigma}(\jmmvec{Q}(t))$. This potential might not be small, but should only be non-zero, cf.
\eref{eq:v_inf}, during a short time interval, given by the time of ``intense interaction'' (collision) of the impinging particle with the surface. Up to first order, the transition rate $p_{ij}$ for an excitation within one spin channel $\sigma$ from an occupied state $\ket{\varepsilon_{i}^{\sigma}}$ into an unoccupied state $\ket{\varepsilon_{j}^{\sigma}}$ (i.e. $i \neq j$), which belong to the part of $h_{(0)}$ that describes the (clean) surface, 
is then given by
\begin{equation}
\label{eq:pij}
  p_{ij}^{\sigma}(t) = \frac{1}{i\hbar} \,
    \bok{\varepsilon_{j}^{\sigma}}{v^{\sigma}(\jmmvec{Q}(t))}{\varepsilon_{i}^{\sigma}} \;
    \exp\left( \frac{i}{\hbar} (\varepsilon_{j}^{\sigma}-\varepsilon_{i}^{\sigma}) \, t \right)
  \; .
\end{equation}
States associated with the impinging particle (spanning the other part of the Hilbert space which $h_{(0)}$ acts on) are not considered here, as the e-h pair excitations to be described are supposed to be located within the substrate. For closer contact with actual calculations, discrete sets of initial and final substrate states rather than continuous band structures are denoted by indices $i$ and $j$, respectively, in the following. In practice, theses indices hence encapsulate $\jmmvec{k}$-point and band indices of the Kohn-Sham states of the clean surface.

\subsection{Electron-hole pair spectra}
\label{ssect:spectra}

The excitation spectrum for a complete round trip is obtained from \eref{eq:pij} by integrating over time and summing over allowed transitions:
\begin{equation}
\label{eq:Pex_1}
\fl \tilde{P}_\jmmid{ex}^{\,\sigma}(\hbar\omega) = 
    \sum_{ij} \left\lvert \, \int\limits_{-\infty}^{+\infty} \rmd t \: p_{ij}(t) \right\rvert^2 \,
              \delta\left(\hbar\omega - (\varepsilon_{j}^{\sigma}-\varepsilon_{i}^{\sigma})\right)
  \; .
\end{equation}
Here $\hbar\omega$ are positive and non-zero excitation energies. Furthermore, here and in the following, for these kind of sums over transitions appropriate weight factors (depending on symmetry) for the $\jmmvec{k}$-point part of the indices $i$ and $j$ are implicitly considered.

Integrating \eref{eq:Pex_1} by parts leads to
\begin{eqnarray}
\label{eq:integration_by_parts}
\fl \int\limits_{-\infty}^{+\infty} \rmd t \: p_{ij}^{\sigma}(t) = \underbrace{  
        \left[ -
        \frac{1}{\varepsilon_{j}^{\sigma}-\varepsilon_{i}^{\sigma}} \:
        \BOK{\varepsilon_{j}^{\sigma}}{\,v^{\sigma}\,}{\varepsilon_{i}^{\sigma}} \;
        \exp\left( \frac{i}{\hbar} (\varepsilon_{j}^{\sigma}-\varepsilon_{i}^{\sigma}) \, t \right) 
        \right]_{t \rightarrow -\infty}^{t \rightarrow +\infty}
      }_{= 0} + \nonumber\\
   \frac{1}{\varepsilon_{j}^{\sigma}-\varepsilon_{i}^{\sigma}} \;
      \underbrace{ 
        \int\limits_{-\infty}^{+\infty} \rmd t \:
        \BOK{\varepsilon_{j}^{\sigma}}{\,\frac{dv^{\sigma}}{dt}\,}{\varepsilon_{i}^{\sigma}} \;
        \exp\left( \frac{i}{\hbar} (\varepsilon_{j}^{\sigma}-\varepsilon_{i}^{\sigma}) \, t \right)
      }_{= \lambda_{ij}^{\sigma}}
  \; .
\end{eqnarray}
Here, the boundary term vanishes due to the properties of the interaction potential $v^{\sigma}$ as given by 
\eref{eq:v_inf}. Reinserting the result of \eref{eq:integration_by_parts} back into \eref{eq:Pex_1} gives
\begin{equation}
\label{eq:Pex_2}
\fl \tilde{P}_\jmmid{ex}^{\,\sigma}(\hbar\omega) = 
    \sum_{ij} \left\lvert 
      \frac{\lambda_{ij}^{\sigma}}{\varepsilon_{j}^{\sigma}-\varepsilon_{i}^{\sigma}} \right\rvert^2
              \delta\left(\hbar\omega - (\varepsilon_{j}^{\sigma}-\varepsilon_{i}^{\sigma})\right)
  \; ,
\end{equation}
where the ``dressed'' (transition) matrix elements
\begin{equation}
\label{eq:lambdaij}
\fl \lambda_{ij}^{\sigma} 
    = \int\limits_{-\infty}^{+\infty} \rmd t \:
      \BOK{\varepsilon_{j}^{\sigma}}{\,\vecgrad_{\jmmvec{Q}} v^{\sigma}(\jmmvec{Q}(t))\,}{\varepsilon_{i}^{\sigma}}
      \boldsymbol{\cdot}
      \jmmvec{\dot{Q}}(t) \; \cdot \;
      \exp\left( \frac{i}{\hbar} (\varepsilon_{j}^{\sigma}-\varepsilon_{i}^{\sigma}) \, t \right)
\end{equation}
are ultimately the key ingredients to be calculated as outlined in \sref{ssect:MEs} below.

Up to now, only occupations of substrate states corresponding to zero temperature have been considered in 
\eref{eq:pij}, \eref{eq:Pex_1} and \eref{eq:Pex_2}. Generalization to finite electronic temperatures is straightforward by introducing Fermi occupation factors
\begin{equation}
\label{eq:Fermi}
  f(\varepsilon) = \frac{1}{\exp\left( \frac{\varepsilon-\varepsilon_\jmmid{F}}{\scakB T} \right) + 1}
  \; .
\end{equation}
Not allowing for de-excitations of ``thermally smeared'' electrons, \eref{eq:Pex_2} then gets extended to the following final working expression for the electron-hole pair spectrum:
\begin{equation}
\label{eq:Pex}
\fl P_\jmmid{ex}^{\,\sigma}(\hbar\omega) = 
    \underbrace{
      \sum_{ij} 
        \left\lvert \frac{\lambda_{ij}}{\varepsilon_{j}^{\sigma}-\varepsilon_{i}^{\sigma}} \right\rvert^2
      \delta\left(\hbar\omega - (\varepsilon_{j}^{\sigma}-\varepsilon_{i}^{\sigma})\right)}
      _{\tilde{P}_\jmmid{ex}^{\,\sigma}(\hbar\omega)} \;
    \left[ f\left(\varepsilon_{i}^{\sigma}\right) - f\left(\varepsilon_{j}^{\sigma}\right) \right] \;
    \theta\left(\hbar\omega\right)
  \; .
\end{equation}
Individual spectra for electrons and holes can be obtained consistently by considering respective transitions only relative to the Fermi level,
\numparts
\begin{eqnarray}
\label{eq:Pex_el}
\fl P_\jmmid{ex,el}^{\,\sigma}(\hbar\omega) & = 
    \sum_{ij} 
      \left\lvert \frac{\lambda_{ij}}{\varepsilon_{j}^{\sigma}-\varepsilon_{i}^{\sigma}} \right\rvert^2
    \delta\left(\hbar\omega - (\varepsilon_{j}^{\sigma}-\varepsilon_\jmmid{F})\right) \;
    \left[ f\left(\varepsilon_{i}^{\sigma}\right) - f\left(\varepsilon_{j}^{\sigma}\right) \right] \;
    \theta\left(\hbar\omega\right) 
  \; , \\
\label{eq:Pex_ho}
\fl P_\jmmid{ex,ho}^{\,\sigma}(\hbar\omega) & = 
    \sum_{ij} 
      \left\lvert \frac{\lambda_{ij}}{\varepsilon_{j}^{\sigma}-\varepsilon_{i}^{\sigma}} \right\rvert^2
    \delta\left(\hbar\omega - (\varepsilon_{i}^{\sigma}-\varepsilon_\jmmid{F})\right) \;
    \left[ f\left(\varepsilon_{i}^{\sigma}\right) - f\left(\varepsilon_{j}^{\sigma}\right) \right] \;
    \theta\left(-\hbar\omega\right)
  \; ,
\end{eqnarray}
\endnumparts
where, as before, the excitation energies $\hbar\omega$ are positive and non-zero. As pointed out by TK \cite{timmer09}
\eref{eq:Pex}, \eref{eq:Pex_el} and \eref{eq:Pex_ho} stay mathematically well defined even for $\left\lvert \varepsilon_{j} - \varepsilon_{i} \right\rvert \rightarrow 0$. Nevertheless, numerical evaluations have difficulties in these regions,
which is why they are omitted in plots of calculated spectra below.

\subsection{Comparison to electronic friction theory}
\label{ssect:EFT}

When electronic friction theory \cite{agliano75,persson82,hellsing84} is combined with the forced oscillator model \cite{trail02,trail03}
and introducing further approximations to obtain excitation spectra \cite{timmer10,mizielinski10}, the equations directly comparable to \eref{eq:Pex_el} and \eref{eq:Pex_ho} read as follows when adapted to the present notation:
\numparts
\begin{eqnarray}
\label{eq:P_FOM_el}
\fl P_\jmmid{FOM; \: el}^{\,\sigma}(\hbar\omega) & = 
    \frac{1}{\pi\hbar}
    \int\limits_{-\infty}^{\varepsilon_\jmmid{F}^{\sigma}} \rmd\varepsilon
      \left\lvert 
        \frac{\lambda_\jmmid{fric}^{\sigma}(\varepsilon_\jmmid{F}^{\sigma} + \hbar\omega - \varepsilon)}
             {(\varepsilon_\jmmid{F} + \hbar\omega) - \varepsilon}
      \right\rvert^{2}
      \left[ f(\varepsilon) - f(\varepsilon_\jmmid{F}^{\sigma} + \hbar\omega) \right] \;
      \theta\left(\hbar\omega\right) \\
\label{eq:P_FOM_ho}
\fl P_\jmmid{FOM; \: ho}^{\,\sigma}(\hbar\omega) & = 
    \frac{1}{\pi\hbar}
    \int\limits_{\varepsilon_\jmmid{F}^{\sigma}}^{+\infty} \rmd\varepsilon
      \left\lvert 
        \frac{\lambda_\jmmid{fric}^{\sigma}(\varepsilon - \varepsilon_\jmmid{F}^{\sigma} + \hbar\omega)}
             {\varepsilon - (\varepsilon_\jmmid{F}^{\sigma} + \hbar\omega)}
      \right\rvert^{2}
      \left[ f(\varepsilon_\jmmid{F}^{\sigma} + \hbar\omega) - f(\varepsilon) \right] \;
      \theta\left(-\hbar\omega\right)
  \; ,
\end{eqnarray}
\endnumparts
with
\begin{equation}
\label{eq:lambda_fric}
  \lambda_\jmmid{fric}^{\sigma}(\varepsilon) = 
    \int\limits_{-\infty}^{\infty} \rmd t \;
      \sqrt{\eta^{\sigma}(Q(t))} \; \dot{Q}(t) \; \exp\left(\frac{i}{\hbar} \varepsilon t \right)
\end{equation}
based on the electronic friction coefficient 
\begin{equation}
\label{eq:eta_electronic}
\fl \eta^{\sigma}(Q(t)) = 
    \pi \hbar \sum_{ij} 
      \left\lvert 
        \BOK{\varepsilon_{j}^{\sigma}(t)} 
            {\frac{\rmd v_\jmmid{eff}^{\sigma}(Q(t))}{\rmd Q}}
            {\varepsilon_{i}^{\sigma}(t)}
      \right\rvert^{2}
    \delta(\varepsilon_{j}^{\sigma}(t) - \varepsilon_\jmmid{F}^{\sigma}) \;
    \delta(\varepsilon_{i}^{\sigma}(t) - \varepsilon_\jmmid{F}^{\sigma}) \;
    \; .
\end{equation}
For the sake of simplicity and better readability, but without loss of generality, a one-dimensional description via a reaction coordinate $Q(t)$ has been employed here.

Comparing \eref{eq:lambdaij} and \eref{eq:lambda_fric} the key ingredient to be calculated for both theories is very similar: Matrix elements of the type
\begin{equation*}
  \BOK{\varepsilon_{j}^{\sigma}(t)}
      {\frac{\rmd v_\jmmid{eff}^{\sigma}(Q(t))}{\rmd Q}}
      {\varepsilon_{i}^{\sigma}(t)}
\end{equation*}
with $v_\jmmid{eff}^{\sigma}(\jmmvec{Q}(t))$ as defined in \eref{eq:vQt}. Obviously, it does not matter for the derivatives in \eref{eq:lambdaij} and \eref{eq:eta_electronic} that the constant part $v_\jmmid{eff; \, non-int}^{\sigma}$, cf. \eref{eq:vQt}, is not subtracted here. In fact, due to the energy differences entering $\lambda_\jmmid{fric}^{\sigma}$ in \eref{eq:P_FOM_el} and \eref{eq:P_FOM_ho}, this can also be read as matrix elements for electronic transitions -- similar to $\lambda_{ij}^{\sigma}$. Their dependence on the electronic structure is less direct though, i.e. ``only'' due to the forced oscillator model connected to the electronic friction coefficient $\eta^{\sigma}$. Additionally, it is important to note, that instantaneous Eigenstates $\ket{\varepsilon_{i}^{\sigma}(t)}$, $\ket{\varepsilon_{j}^{\sigma}(t)}$ are used to obtain the latter. Obviously, the present theory can in principle be modified in a straightforward way to use instantaneous Eigenstates as well in \eref{eq:lambdaij}. However, this leads to conceptual issues with shifts of Eigenvalues in the DFT calculations at different points of a trajectory and technical problems with (relative) phase shifts of the instantaneous Eigenstates.

Apart from this, the main difference between the present and electronic friction theory comes from the fact that squared moduli of the (respective) matrix elements appear before the time integration is carried out in \eref{eq:lambda_fric} and \eref{eq:eta_electronic}. As discussed extensively by TK \cite{timmer09}, this is the reason why \eref{eq:lambdaij}, \eref{eq:Pex}, \eref{eq:Pex_el} and \eref{eq:Pex_ho} remain mathematically well defined even in case of an adiabatic spin transition: The spin population assigned to the impinging adsorbate typically (at least for hydrogen atoms impinging on different substrates)
\cite{lindenblatt06a,lindenblatt06b,lindenblatt06c,trail02,trail03,timmer09} is proportional to $\left\lvert \jmmvec{Q}(t) - \jmmvec{Q}_{0} \right\rvert^{1/2}$ around the transition point $\jmmvec{Q}_{0}$. This leads to a $\left\lvert \jmmvec{Q}(t) - \jmmvec{Q}_{0} \right\rvert^{-1/2}$ singularity of ${\rmd v_\jmmid{eff}^{\sigma}(Q(t))}/{\rmd Q}$, for which the aforementioned main difference is of crucial (mathematical) importance.

\subsection{Dissipated energy}
\label{ssect:Ediss}

Finally, the energy dissipated into electron-hole pair excitations, which is the key objective here, is obtained as energy-weighted integral over the excitation spectrum, \eref{eq:Pex}:
\begin{equation}
\label{eq:Eeh}
  E_\jmmid{eh}^\jmmid{\,\sigma} = 
    \int\limits_{0}^{+\infty} \rmd\varepsilon \; \varepsilon \, P_\jmmid{ex}^{\,\sigma}(\varepsilon)
  \; . \\
\end{equation}
Again, a comparable expression is also obtained within electronic friction theory,
\begin{equation}
\label{eq:Eeh_fric}
  E_\jmmid{eh;\: fric}^\jmmid{\,\sigma} = 
    \int_{\jmmvec{Q}} \rmd Q \; \eta^{\sigma}  =   \int\limits_{\underline{\tau} \rightarrow -\infty}
               ^{\overline{\tau} \rightarrow +\infty} 
      \rmd t \; \eta^{\sigma}(\jmmvec{Q}(t)) \; \dot{\jmmvec{Q}}(t)
  \; ,
\end{equation}
which additionally would allow to ``measure'' the dissipated energy for only a part of the considered trajectory $\jmmvec{Q}(t)$. However, with friction theory not being (directly) applicable here, \eref{eq:Eeh} should serve well for providing the desired estimate of energy dissipated into electron-hole pairs.

\subsection{Realistic first-principles trajectories}
\label{ssect:traj}

In previous work addressing e-h pair excitation typically strictly one-dimensional model trajectories $Q(t)$ along the surface normal have been considered when studying adsorption of mono-atomic adsorbates with perpendicular impingement over high symmetry sites \cite{lindenblatt06a,lindenblatt06b,lindenblatt06c,trail02,trail03,timmer09,timmer10,mizielinski10}. The motion of incoming poly-atomic particles (like the diatomic molecule O$_2$) is in contrast less likely to be reducible to a single spatial dimension. To keep the input data entirely scalar, an effectively one-dimensional description of coordinates and velocities, i.e. a suitable reaction coordinate $Q$, must therefore be employed. A canonical choice is the arc length associated with a given trajectory $\jmmvec{Q}(t)$:
\begin{equation}
\label{eq:Qarc}
  Q(t) = \int\limits_{\underline{\tau}}^{t} \rmd\tau \; \lVert \Delta \jmmvec{Q}(\tau) \rVert_{2}
  \quad .
\end{equation}
However, we note that in case of non-negligible changes of relative coordinates within the adsorbate, \eref{eq:Qarc} does not provide immediately obvious physically meaningful information about the trajectory of an adsorbate. For a stiff diatomic molecule like O$_2$ this might not be much of a concern, unless considering large vibrational excitations e.g. during scattering at high kinetic energies. Still, it is interesting to note that  \eref{eq:lambdaij} and all its descendants remain invariant, if trajectories are described by scaled and shifted coordinates and consistent velocities
\numparts
\begin{eqnarray}
\label{eq:QtransQ}
  \jmmvec{Q}(t) & \quad\longrightarrow\quad 
    \jmmvec{\tilde{Q}}(t) \equiv \jmmvec{Q}_{0} + \alpha \, \jmmvec{Q}(t) \\[0.1em]
\label{eq:QtransQdot}
  \jmmvec{\dot{Q}}(t) & \quad\longrightarrow\quad
    \jmmvec{\dot{\tilde{Q}}}(t) \equiv \alpha \, \jmmvec{\dot{Q}}(t)
  \quad .
\end{eqnarray}
\endnumparts
When $\alpha = 1 / \sqrt{N_\jmmid{ads}}$ for an adsorbate consisting of ${N_\jmmid{ads}}$ atoms, the corresponding arc length $\tilde{Q}(t)$ then conveniently corresponds to the absolute distance travelled by the centre of mass of the adsorbate (or likewise any of its constituents).

The trajectories employed in the present work have been obtained from molecular dynamics simulations on a six-dimensional (Born-Oppenheimer) potential energy surface (PES) $V_{\rm 6D}$, describing the molecular degrees of freedom of an isolated O$_2$ molecule on a frozen Pd(100) surface. This continuous PES representation is constructed by a highly accurate interpolation of discrete DFT energies computed within the Perdew, Burke and Ernzerhof (PBE) generalized gradient approximation to electronic exchange and correlation \cite{perdew96,perdew97}. Further details and a statistical analysis of the resulting dissociation dynamics will be published elsewhere \cite{meyer11a}. Briefly, the interpolation is based on neural networks, adapted to properly take symmetry into account \cite{meyer11b}. 4000 single-point DFT calculations in supercell geometries, carefully sampling molecular positions and orientations over the Pd(100) substrate, serve as database for the interpolation. The supercell geometry is characterized by $(3\times3)$ supercells, a five layer slab, and a vacuum distance of 15\,{\AA}. A $(4\times4\times1)$ Monkhorst-Pack grid is used for $\jmmvec{k}$-point sampling. Electronic states are described by a plane wave basis using a cut-off energy of 400 eV together with ultrasoft pseudopotentials as implemented in the \CASTEP\ code and supplied within the default \CASTEP\ pseudopotential library \cite{clark09}. 

\subsection{Efficient matrix element calculation}
\label{ssect:MEs}

The electronic structure data obtained concomitantly with the single-point energetics for the PES then also forms the basis for the evaluation of the matrix elements entering \eref{eq:lambdaij}. In several applications of electronic friction theory, the corresponding matrix elements given by \eref{eq:lambda_fric} are related to the nuclear kinetic energy terms, which are neglected within the Born-Oppenheimer approximation. Following ideas from Head-Gordon and Tully \cite{gordon92,gordon95}, the commonly used plane wave implementation by Lorente and Persson \cite{lorente00a,lorente00b,teobaldi07} makes use of first order changes in Kohn-Sham states and Eigenvalues to obtain the aforementioned matrix elements. Obviously, this cannot work in the present context when dealing with unperturbed states of the substrate. In the original TK work \cite{timmer09} and in an equivalent strategy previously used within electronic friction theory \cite{trail01}, the matrix elements are calculated directly through the derivative of the effective Kohn-Sham potential. Quite in contrast, here we follow a different strategy instead, namely to compute $M_{ij}^\sigma = \BOK{\varepsilon_{j}^{\sigma}}{\, v_\jmmid{eff}^{\sigma}(Q) \,}{\varepsilon_{i}^{\sigma}}$ at all points $Q(t_{n})$ for which electronic structure data is available. Afterwards, each of these matrix elements is interpolated along $Q$ and the analytical derivative of the interpolation gives
$\partial M_{ij}^{\sigma}(Q) / \partial Q$. Through
\begin{equation}
\label{eq:dVijdQdQdt_JM}
  \BOK{\varepsilon_{j}^{\sigma}}
      {\, \frac{\rmd v^{\sigma}(\jmmvec{Q}(t))}{\rmd t} \,}
      {\varepsilon_{i}^{\sigma}} 
  \: \approx \:
  \frac{\partial}{\partial Q} M_{ij}^{\sigma}(Q)
    \cdot
    \dot{Q}(t)
  \quad ,
\end{equation}
this then also yields the matrix elements required to determine the e-h spectrum via \eref{eq:lambdaij}. Following common practise to rely on sufficient decoupling between periodic images \cite{lindenblatt06a,lindenblatt06b,lindenblatt06c,timmer09,lorente00a,lorente00b,teobaldi07}, only \emph{intra}-$\jmmvec{k}$ transitions are considered in this procedure, which becomes exact for a periodic overlayer \cite{trail01}.

The crucial advantage of this new strategy is that it allows for efficient use of existing infrastructure in most plane wave codes, because the derivative of the potential does not need to be constructed in a form which can be used to act on the states directly. This is particularly useful (if not essential) when dealing with non-local and/or ultrasoft pseudopotentials. Notwithstanding, the matrix elements calculated for all $Q(t_{n})$ have to be kept in memory to perform the interpolation and the ensuing derivative. This does lead to higher memory demands than for other schemes, which by virtue of (memory) parallelisation nevertheless does no longer constitute an invincible problem. We have carefully validated and benchmarked our implementation strategy by recomputing the H at Al(111) data reported in the original TK work \cite{timmer09}. Quantitative agreement is reached at a minute fraction of the computational cost. Similar observations have also been made by Timmer and Kratzer in their more recent work \cite{timmer10} after also adapting to our implementation strategy.

In theory, the loss of orthonormality between substrate states when using ultrasoft pseudopotentials
\cite{vanderbilt90,laasonen93} could pose a serious problem for \eref{eq:pij}, hence shaking the foundations of the perturbative approach. However, in practice, the aforementioned tests for H at Al(111) have shown that this is not a general concern. Selected comparisons of the results for the present system to results obtained with norm-conserving pseudopotentials indicate some quantitative dependence of the ultimately deduced dissipated energy, yet, without touching on any of the physical conclusions discussed below. Much more problematic is the well-known spurious ferromagnetism of Pd within semi-local DFT \cite{alexandre06a,alexandre06b}. Test calculations indicate that this leads to a dramatic increase of the dissipated energy that can be of the order of several 100\,\% in relative terms. For all results presented below, matrix elements have therefore been evaluated using the Kohn-Sham states of a non-spin polarized clean Pd(100) surface that have been ``cloned'' into both spin channels of the (spin-polarized) adsorption calculations including the O$_2$ molecule. In this respect the fact that only the derivative of the potential enters \eref{eq:lambdaij} can be seen as particular (additional) virtue of the employed scheme for this specific system: It allows to exploit a cancellation of errors in differences (as already in case of the energetics) -- quite in contrast to a conventional treatment e.g. within time-dependent DFT.

\section{Results}
\label{sect:results}

\subsection{Electron-Hole Pair Spectra and Dissipated Energies}
\label{ssect:spec_Ediss}

\begin{figure}
\centering
\includegraphics{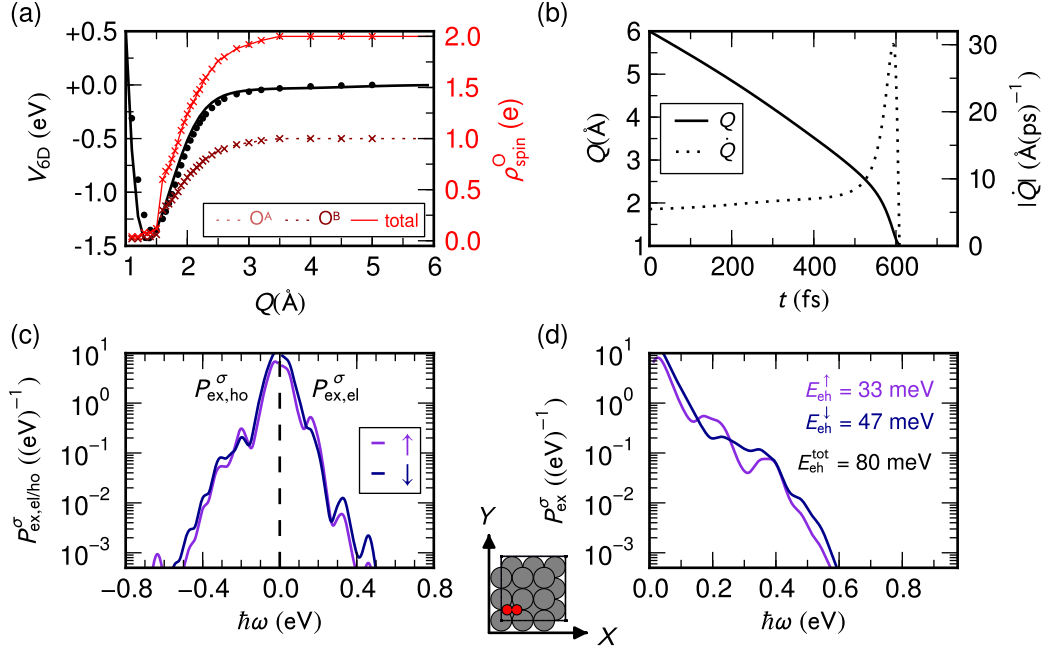}
\caption[Electron-hole pair excitations for an O$_2$ molecule impinging side-on over a hollow site of Pd(100).]
{Electron-hole pair excitations created by an O$_2$ molecule impinging side-on above a hollow site ({\em h-para}) as shown in the inset at the bottom. (a) PES $V_\jmmid{6D}$ along the trajectory given by the reaction coordinate $Q$ (neural network interpolation = black solid line, DFT input data = black circles), as well as projections of the spin density onto the two constituting oxygen atoms (O$^A$, O$^B$ = dotted lines in shades of dark red, sum of O$^A$ and O$^B$ = light red solid line). (b) Evolution of reaction coordinate $Q(t)$ and corresponding velocity $\dot{Q}(t)$ with time $t$ along the trajectory. (c) Separate electron (at positive excitation energies) and hole (at negative excitation energies $\hbar{\omega}$) spectra $P_\jmmid{ex,el}^{\,\sigma}(\hbar\omega)$ and $P_\jmmid{ex,ho}^{\,\sigma}(\hbar\omega)$ according to \eref{eq:Pex_ho} and \eref{eq:Pex_el}. (d) Total e-h pair spectrum $P_\jmmid{ex}^{\,\sigma}(\hbar\omega)$ given by \eref{eq:Pex} together with resulting dissipated energies according to \eref{eq:Eeh}. All spectra are for a half round trip with energies $\hbar\omega$ relative to the Fermi energy. Both majority ($\uparrow$, violet) and minority ($\downarrow$, blue) spin channels are shown.}
\label{fig:hpara}
\end{figure}

\begin{figure}
\centering
\includegraphics{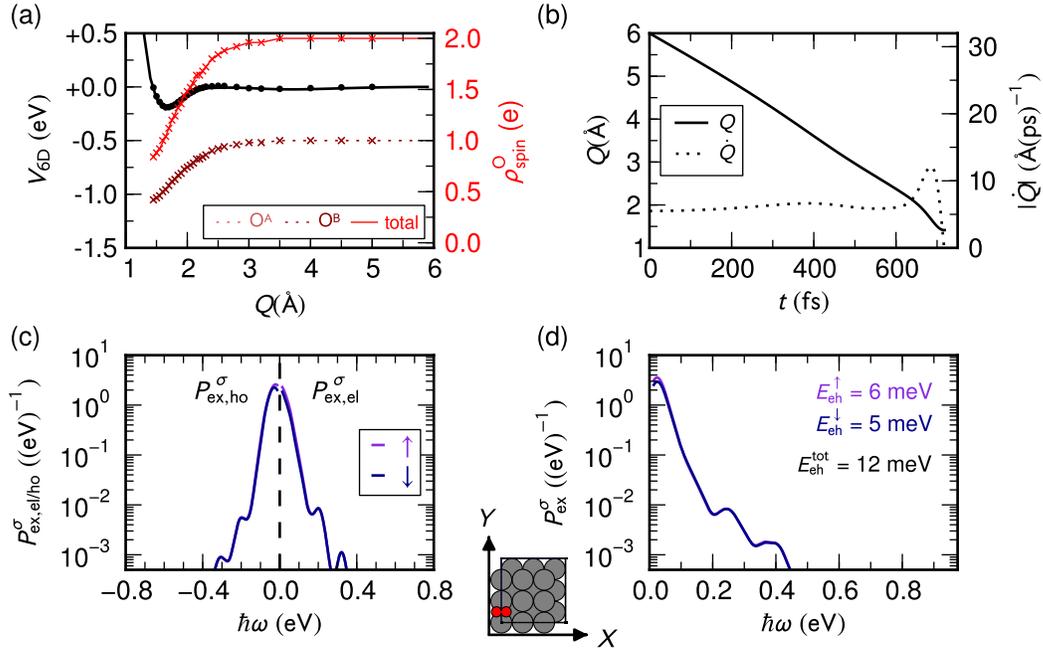}
\caption[Electron-hole pair excitations for an O$_2$ molecule impinging side-on over a bridge site of Pd(100).]
{Same as \fref{fig:hpara}, but for an O$_2$ molecule impinging side-on above a bridge site ({\em b-para}) as shown in the inset at the bottom.}
\label{fig:bpara}
\end{figure}

\begin{figure}
\centering
\includegraphics{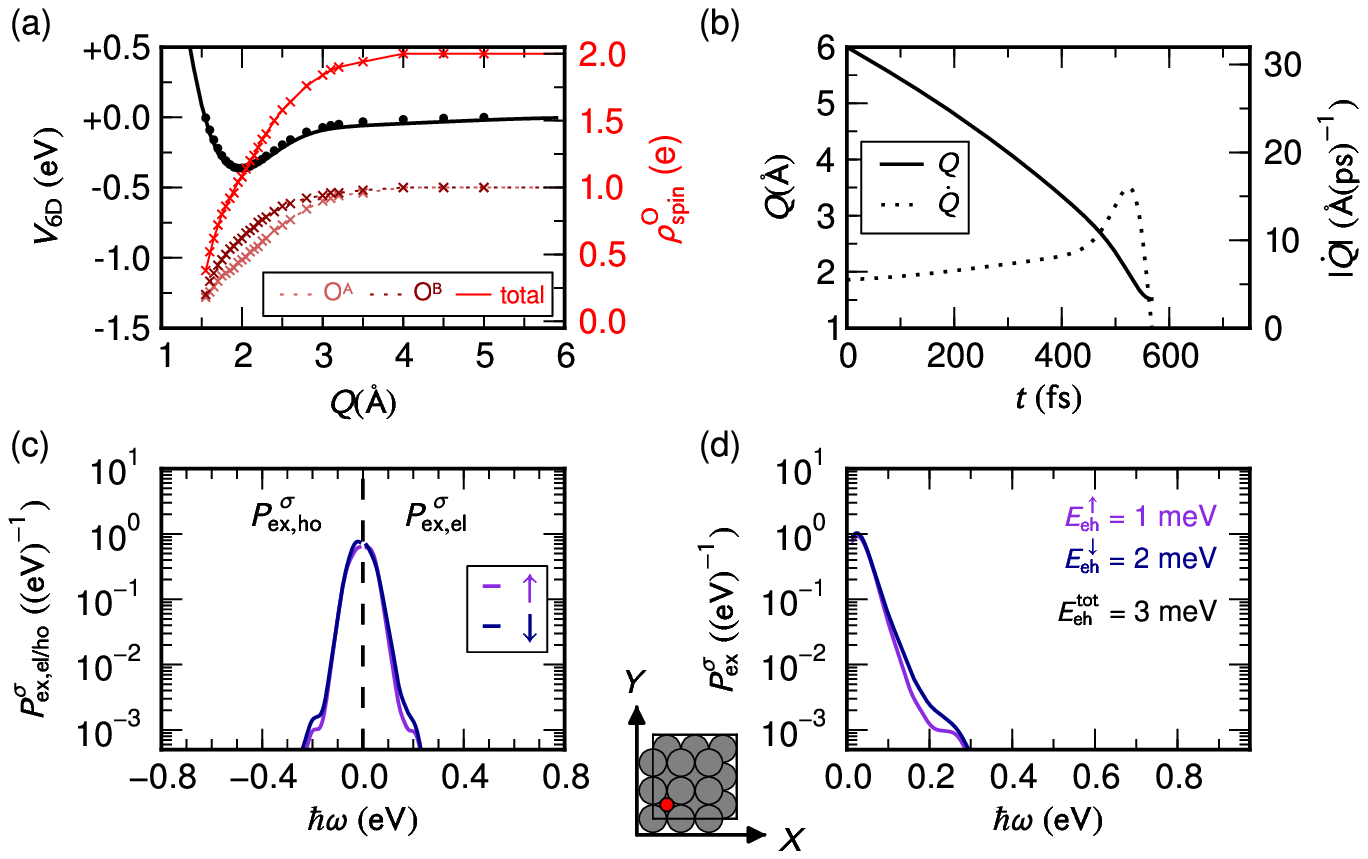}
\caption[Electron-hole pair excitations for an O$_2$ molecule impinging head-on over a hollow site of Pd(100).]
{Same as \fref{fig:hpara}, but for an O$_2$ molecule impinging head-on above a hollow site ({\em h-perp}) as shown in the inset at the bottom.}
\label{fig:hperp}
\end{figure}

\begin{figure}
\centering
\includegraphics{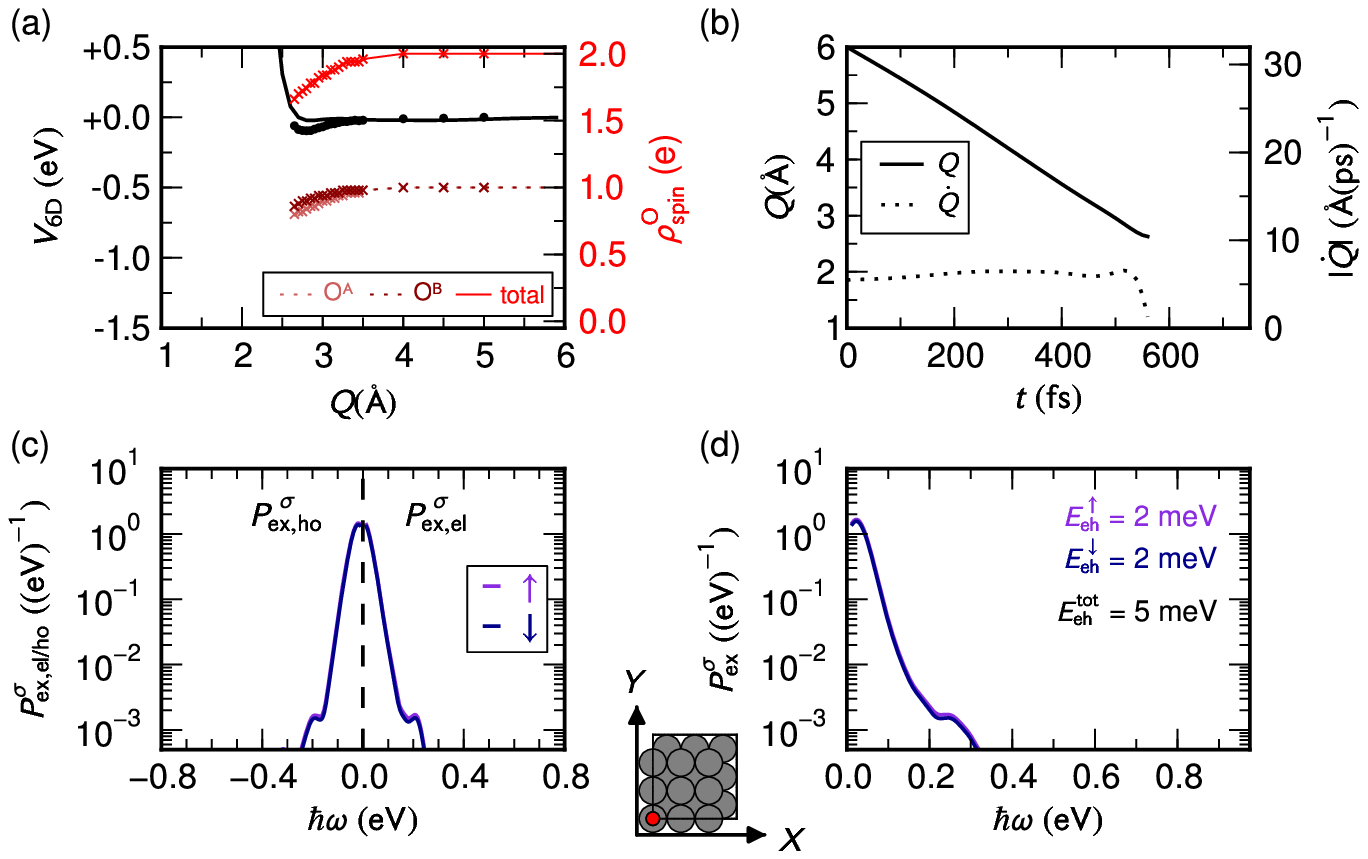}
\caption[Electron-hole pair excitations for an O$_2$ molecule impinging head-on over a top site of Pd(100).]
{Same as \fref{fig:hpara}, but for an O$_2$ molecule impinging head-on above a top site ({\em t-perp}) as shown in the inset at the bottom.}
\label{fig:tperp}
\end{figure}

With the objective to compute e-h pair spectra and to estimate the
concomitant energy dissipated into electronic excitations we focus our
investigation on four distinct trajectories, selected to span the space of
possible O$_2$ impingement over the Pd(100) surface.  Namely these are two
trajectories in which the O$_2$ molecule initially approaches with its
molecular axis parallel to the surface and oriented along a [100] direction
(a so-called side-on configuration), and two trajectories in which the O$_2$
molecule initially approaches with its molecular axis perpendicular to the
surface (so-called head-on configuration).  In order to sample the influence
of the lateral corrugation of the surface potential one of the two side-on
trajectories starts out with the O$_2$ center of mass located above a
fourfold hollow site, and one trajectory with this centre above a twofold
bridge site, cf.  insets in figures~\ref{fig:hpara} and \ref{fig:bpara}. 
Similarly the two head-on trajectories start out above a hollow and above a
top site, cf.  insets in figures~\ref{fig:hperp} and \ref{fig:tperp}. 
Henceforth we will use the intuitive abbreviations {\em h-para}, {\em
b-para}, {\em h-perp} and {\em t-perp} for the four cases, respectively. 
Each trajectory begins with the O$_2$ center of mass at a perpendicular
distance $Z_{0} = 6$\,{\AA} above the surface, where the adsorbate-substrate
interaction potential, i.e.  the PES $V_{\rm 6D}$, has already well decayed
to zero.  As an initial kinetic energy we choose 50\,meV to mimic thermal
molecules in the high-energy Boltzmann tail.  
In general, non-adiabatic energy losses should increase with higher velocities,
which is why we expect the deduced dissipated energy to represent a useful 
upper bound for a thermal gas.
However, in order to explicitly test the dependence on
the initial velocity quantitatively we additionally consider a {\em h-para}
trajectory with an increased initial kinetic energy of 400\,meV,
corresponding to an almost tripled initial velocity from $\dot{Q}(0) =
5.5$\,{\AA}/fs to $\dot{Q}(0) = 15$\,{\AA}/fs.

The results for the four trajectories {\em h-para}, {\em b-para}, {\em h-perp} and {\em t-perp} with low initial kinetic energy are compiled in figures~\ref{fig:hpara} to \ref{fig:tperp}, respectively. In all four cases the molecule is reflected, such that a half round trip as assumed in the perturbative approach can be consistently defined. With the definition given in \eref{eq:QtransQ} the reaction coordinate $Q(t)$ corresponds for the considered trajectories essentially to the vertical distance $Z(t)$ of the O$_2$ center of mass above the surface. As shown in parts (b) of figures~\ref{fig:hpara} -- \ref{fig:tperp} the half round trip proceeds in all cases in about 600 fs, in which the molecule approaches from its initial height of $Q(0) \approx Z(0) = 6.0$\,{\AA} towards the surface. Intuitive in terms of the expected corrugation of the surface potential, the point of reflection (defining the end of the considered half round trip) is with $Z \approx 2.6$\,{\AA} highest for the {\em t-perp} trajectory with the molecule head-on above the top site, and with $Z \approx 1.0$\,{\AA} lowest for the {\em h-para} trajectory with the molecule side-on above the hollow site. In the other two trajectories the reflection occurs at intermediate $Z$ heights. 

These differences in the point of closest encounter have direct consequences on the adiabatic spin transition of the O$_2$ molecule. As illustrated by the projected O$_2$ spin density in parts (a) of figures~\ref{fig:hpara} -- \ref{fig:tperp} the molecule approaches the surface only sufficiently closely in the {\em h-para} trajectory for the spin to be completely quenched at the point of reflection. In this case exactly the square-root like decay of the spin density results that has been generally observed for hydrogen adsorption before \cite{lindenblatt06a,lindenblatt06b,lindenblatt06c,trail02,trail03,timmer09} and which as described in \sref{ssect:EFT} would prevent application of electronic friction theory. In contrast, in the {\em t-perp} trajectory the transition to the singlet state has not even really started when the molecule gets reflected. However, the adiabatic spin transition of the oxygen molecule depends not only on the penetration depth, but also strongly on lateral position and orientation. This is nicely illustrated by the trajectories {\em b-para} and {\em h-perp} in figures~\ref{fig:bpara}(a) and \ref{fig:hperp}(a), respectively. In both cases the minimum $Z$-height reached is comparable, but the projected spin density on both oxygen atoms remains nearly twice as large for {\em b-para}.

Significant acceleration beyond the initial velocity of $\dot{Q}(0) = 5.5$\,{\AA}/fs only takes place in areas of attractive potential at low $Z$ heights just before the molecule hits the repulsive Pauli wall. As shown in \fref{fig:hpara}(b) this is obviously largest for the most favourable {\em h-para} configuration, which is very close to the main entrance channel for dissociation in this system \cite{meyer11a}. While the maximum speed reached is in this case about five times the initial velocity, essentially no acceleration occurs throughout the entire trajectory in the most repulsive {\em t-perp} case, cf. \fref{fig:tperp}(b). These differences have to zeroth order exactly the bearings on the generated e-h pairs and concomitant energy dissipation one would expect e.g. from electronic friction theory. As shown both separately for electrons and holes in figures \ref{fig:hpara}(c) --\ref{fig:tperp}(c) and combined into one e-h pair spectrum in figures~\ref{fig:hpara}(d) -- \ref{fig:tperp}(d), the qualitative shape of the spectra is for all trajectories about the same. However, the absolute magnitude is much larger for the high velocity attractive case {\em h-para}, resulting into a total amount of dissipated energy that is about an order of magnitude larger than in the other three cases. Notwithstanding, the latter three cases illustrate that at closer look also other factors like the orientation of the molecule matter: Specifically in the trajectories {\em b-para} and {\em h-perp} in figures~\ref{fig:bpara}(a) and \ref{fig:hperp}(a), respectively, the O$_2$ molecule experiences the mentioned similarly repulsive interaction potential and reaches comparable maximum speed before reflection. Nevertheless, the energy dissipation into e-h pairs is by a factor of four less for the latter. This constitutes a nice confirmation of one of the key results of the electron friction work of Juaristi \etal \cite{juaristi08}: The importance of the ``high'' dimensionality of the molecule-substrate-interaction also extends to e-h pair excitations, such that a proper assessment of the role of this dissipation channel needs necessarily to rely on a representative set of impingement scenarios, not just one model trajectory.

Further insight into the velocity dependence comes from the final {\em h-para} trajectory, in which the initial kinetic energy has been raised to 400\,meV. Despite the nearly tripled initial velocity, the corresponding trajectory and spectra (not shown) do not change qualitatively with respect to the data shown in \fref{fig:hpara}. The total energy dissipated into e-h pairs is increased by around a factor of 1.3, resulting in a total amount of about 100\,meV. Compared to the electronic friction theory results for N$_{2}$ on W(110) \cite{juaristi08}, the relative increase of the total dissipated energy is thus much less. There, for perpendicular incidence, a similar increase of the incidence kinetic energy has led on average to an increase in the energy dissipation of more than an order of magnitude. At present it is unclear whether these differences are due to the two different systems studied, or whether they indicate potential shortcomings of either theory. Direct application of the present theory to the N$_2$ on W(110) system will therefore constitute an interesting next step in our work.

Among the trajectories considered, the total dissipated energy per half round trip is thus largest for exactly the {\em h-para} trajectory that is ``most interesting'' for thermal impinging molecules in the sense that it falls very close to the dominant entrance channel to dissociation into which almost all molecules get steered \cite{meyer11a}. For this {\em h-para} trajectory the total dissipated energy furthermore seems not very sensitive to the specific initial velocity. Considering also the approximate nature of the employed perturbative approach we therefore take $\sim 100$\,meV as a reasonable estimate of the order of magnitude of the electronic dissipation channel. With the present DFT-PBE setup we compute a total adsorption energy released in the adsorption of O$_2$, i.e. in the exothermic dissociation into two separate O atoms at the Pd(100) surface, of 2.6\,eV. As such the e-h pair excitation channel does not even take up 5\,\% of this energy, which provides an {\em a posteriori} justification for the assumption of a weak perturbation underlying the present approach. On the one hand, the insignificance of this channel for this system is somewhat consistent with the inability to detect chemicurrents during the adsorption of O$_2$ on polycrystalline palladium \cite{nienhaus09}. On the other hand, it is remarkably low compared to H impingement over the threefold  hollow site of Al(111). The latter leads to a comparable release of chemisorption energy, but the computed amount of energy taken up by e-h pairs was one order of magnitude more, i.e. 1\,eV (albeit likely favoured by a penetration of the adsorbate into the first substrate layer) \cite{lindenblatt06b}. Neither is it thus possible to {\em a priori} dismiss electronic excitations in the adsorption process at metal surfaces, nor does the difference between the two systems follow the picture expected from the much higher Fermi-level DOS in the Pd(100) system.

\subsection{Spin transition}

At a closer look at the e-h pair spectrum for the {\em h-para} trajectory shown in \fref{fig:hpara}(d) a slight asymmetry between electrons and holes deserves further attention. The fact that excitations are stronger in the spin (down) minority channel, and specifically in the high-energy wing of the spectrum, is particularly interesting in light of the spin transition of the O$_2$ molecule, which, according to \fref{fig:hpara}(a), is completed for this trajectory before the point of reflection is reached. A similar asymmetry, albeit much weaker, also seems to be present for the {\em h-perp} trajectory, cf. \fref{fig:hperp}, correlating with the aforementioned similar degree of spin quenching in this case. Timmer and Kratzer have obtained nothing comparable in the e-h pair spectra of H on Al(111) \cite{timmer09}. In contrast, for the same system Lindenblatt and Pehlke \cite{lindenblatt06a} did determine a similar abundance of excited electrons and holes in the spin majority and minority channel, respectively. However, since the time-dependent DFT based dynamics is intrinsically non-adiabatic and the spectra are obtained {\em a posteriori} by relaxation back to the electronic ground state, this is not surprising. It can be easily understood by propagation on excited PESs during the non-instantaneous spin transition. In contrast, the present description is adiabatic with respect to the spin transition, and e-h pair excitations are evaluated based on unperturbed substrate states. 

\begin{figure}
\centering
\includegraphics{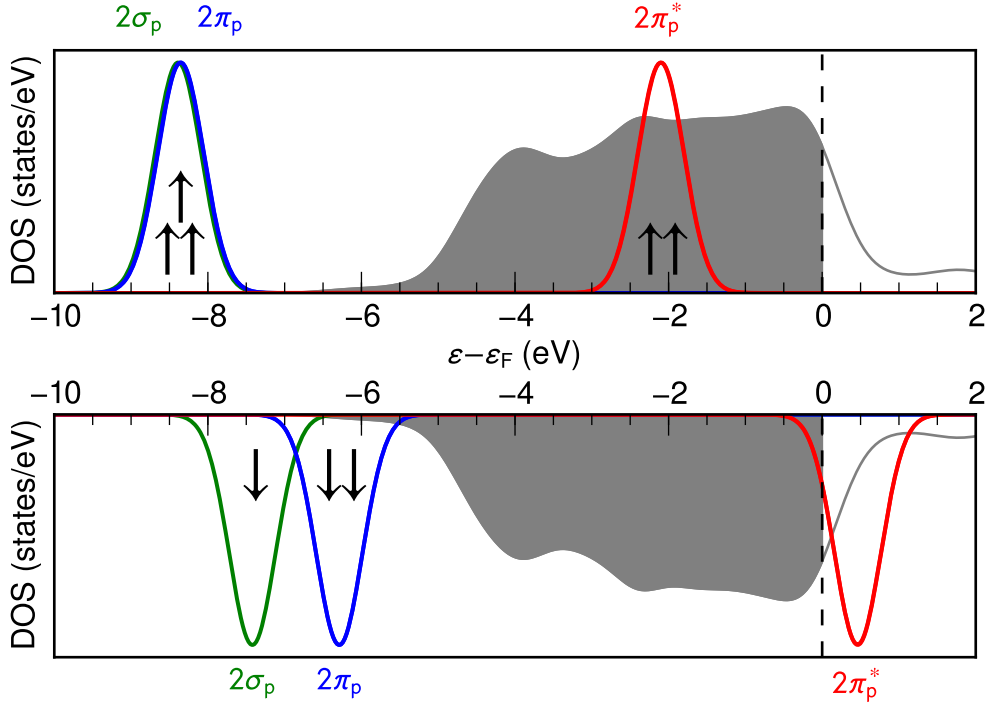}
\caption[Density of states including projections onto molecular orbitals at the beginning of the {\em h-para} trajectory.]
{DOS including projections onto high-lying O$_2$ molecular orbitals at the beginning ($t=0$) of the {\em h-para} trajectory, when there is still negligible interaction with the Pd(100) substrate. The upper panel shows the majority spin channel (spin $\uparrow$), the lower panel shows the minority spin channel (spin $\downarrow$). Note that at this large molecule-surface distance the O$_2$ molecular states $2\sigma$, $2\pi$ and $2\pi^*$ are still rather sharp and are here only broadened by a Gaussian with a width of 0.3\,eV for better visualization. The DOS of the clean Pd(100) surface (downscaled by a factor of 50) is shown in black, with the occupation of states below the Fermi level $\varepsilon_{\rm F}$ represented through the grey filling.}
\label{fig:PDOS}
\end{figure}

\begin{figure}
\centering
\includegraphics{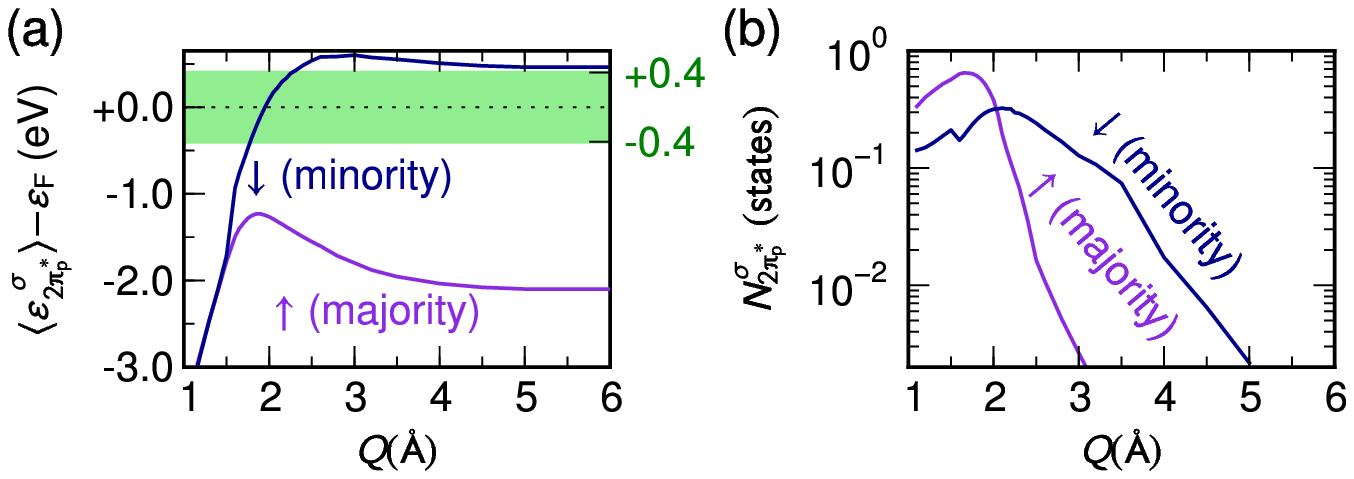}
\vspace{2em}
\caption[Change of O$_2$ projected states along the \emph{h-para} trajectory.]
{(a) Evolution of the center of the ${2\pi^{*}}^{\sigma}$ molecular orbital $ \langle \varepsilon_{2\pi^{*}}^{\sigma}(Q) \rangle $
along the {\em h-para} trajectory. (b) Corresponding number of states projected onto this orbital in an energy interval 0.4\,eV around the Fermi-level, i.e. around the dominant energy range of e-h pair excitations highlighted by the green area in panel (a).}
\label{fig:O2states}
\end{figure}

In order to put the observed spin asymmetry into perspective with the spin transition,
a detailed analysis of the O$_2$ molecular states interacting with the Pd(100) surface 
along the trajectory $Q(t)$ can be illustrative.
For this we evaluate the density of states projected onto a Kohn-Sham state $\ket{\varphi_\jmmid{O_2}^{\sigma}(Q)}$ of spin $\sigma$ of the free O$_2$ molecule (molecular PDOS) \cite{meyer11_2}
\begin{equation}
\label{eq:PDOS}
  \rho_{\varphi_\jmmid{O_2}}^{\sigma}(\varepsilon;Q) = \sum_n
    \jmmabs{ \BraKet{\varphi_\jmmid{O_2}^{\sigma}(Q)}{\varepsilon_n^{\sigma}(Q)} }^2
    \delta\left(\varepsilon - \varepsilon_n\right)
  \quad .
\end{equation}
It is important to emphasize that here the Kohn-Sham states of the interacting system
of substrate and molecule $\ket{\varepsilon_n^{\sigma}(Q)}$ are used for this analysis.
The most relevant state with respect to overlap with the Pd valence $d$-band is specifically the degenerate $2\pi^*$ state \cite{behler05}, which in the free triplet O$_2$ molecule is occupied by two electrons in the spin majority channel and unoccupied in the spin minority channel. Upon approaching the surface, the interaction with the substrate potential starts to broaden the discrete molecular levels. Their centre of energy along the trajectory $Q(t)$ is each given by
\begin{equation}
\label{eq:PDOS_peaks}
  \langle \varepsilon_{2\pi^{*}}^{\sigma} \rangle(Q) = 
    \int_{-\infty}^{\infty} 
      \rmd\varepsilon \; \varepsilon \; \rho_{2\pi^{*}}^{\sigma}(\varepsilon;Q)
  \quad .
\end{equation}
As shown in \fref{fig:PDOS}, at the beginning of the calculated trajectory this centre is for the majority $2\pi^{*\uparrow}$ level located about 2\,eV below the Fermi-level, while it is at about 0.5\,eV above the Fermi-level for the minority $2\pi^{*\downarrow}$ level.
If from inspection of \fref{fig:hpara}(d) we roughly take the dominant fraction of e-h pair excitations to spread over a range of 0.4\,eV around the Fermi-level, we see that it is primarily the spin minority $2\pi^{*\downarrow}$ level that falls close to this range. To quantify this as the width of the $2 \pi^*$ level increases along the trajectory, we evaluate the number of projected states that lie within this relevant energy range as
\begin{equation}
\label{eq:PDOS_states}
  N_{2\pi^{*}}^{\sigma}(Q) = 
    \int_{\varepsilon_\jmmid{F}-0.4{\rm eV}}^{\varepsilon_\jmmid{F}+0.4{\rm eV}} 
      \rmd\varepsilon \; \rho_{2\pi^{*}}^{\sigma}(\varepsilon;Q) \quad .
\end{equation}
\Fref{fig:O2states} displays this quantity together with the evolution of the centre of the $2\pi^*$ level in both spin channels along the {\em h-para} trajectory. A much larger number of corresponding projected states is clearly visible for the spin minority channel specifically in the range of vertical heights $\sim 2-5$\,{\AA} that is most relevant for the spin transition. Qualitatively the same result is also obtained when reducing the considered energy range around the Fermi-level from 0.4\,eV to lower values and thereby concentrating on the more and more dominant part of the obtained e-h pair spectrum, cf. \fref{fig:hpara}. 

Altogether we therefore conclude that the observed stronger e-h pair excitations in the minority spin channel coincide with a larger amount of substrate states in the corresponding energy range, which overlap with the $2\pi^{*\downarrow}$ orbital of the O$_2$ molecule. A tempting interpretation is therefore to see the excess excitations as a reflection of the preferred tunneling of excited minority spin electrons that leads to the quenching of the oxygen spin-density in the here assumed adiabatic spin transition picture. If the electronic structure during the ``real'' spin transition does not behave in a radically different fashion, this connection between the different energetic position of majority and minority O$_2$ $2\pi^*$ level and corresponding different overlap with the energy range spanned by e-h pair excitations would prevail. In that case, this then provides an intuitive mechanism how electron tunneling from the (here forcibly) non-spin polarized substrate electronic structure compensates the adsorbate spin.

\section{Conclusions}
\label{sect:conlusions}

In conclusion we have advanced and employed a first-principles perturbative approach to study the energy dissipation into e-h pair excitations during the adsorption of O$_2$ at Pd(100). Despite the unusually large Pd density of states at the Fermi-level, which in principle could render this system a prime candidate for electronic non-adiabaticity, we obtain energy losses into this electronic channel that are at maximum of the order of 5\,\% of the total released adsorption energy. In detail, this loss and the underlying electron-hole pair spectra differ considerably for different O$_2$ impingement configurations, i.e. they depend on the molecular orientation with respect to the surface and on the lateral position of impact. By far the largest amount of excitations is obtained for a side-on approach above the fourfold hollow site. This is also the statistically most relevant configuration as it corresponds closely to the entrance channel for dissociation into which the dominant fraction of impinging molecules gets steered. While these variations confirm the necessity to account for the ``high'' dimensionality of the molecule-substrate-interaction by considering appropriate statistical averages over ensembles of trajectories, we find only a weak dependence of the total dissipated energy into e-h pairs on the initial O$_2$ velocity. 

All in all we therefore derive from our calculations a value of $\sim 100$\,meV as a reasonable estimate of the order of magnitude for the electronic dissipation channel in this system. Even when cautiously considering that the approximations underlying the employed perturbative approach lead to a similar underestimation of this quantity as was identified for H at Al(111) \cite{timmer09}, e-h pair excitation thus still plays an insignificant role for the dissipation of the large amount of 2.6\,eV adsorption energy released during dissociative adsorption. This assessment is very much in line with the trends observed in other studies beyond single atoms at metal surfaces, no matter if the impinging diatomic molecule carries a permanent dipole moment (HCl on Al(111) \cite{pehlke10}) or not (H$_2$ on Cu(110), N$_2$ on W(110) \cite{juaristi08}), as well as with the hitherto unsuccessful attempts to detect chemicurrents in experiments over polycrystalline palladium \cite{nienhaus09}.

Notwithstanding, two aspects underlying the present state-of-the-art treatment need to be emphasized that could in principle radically question our conclusions on the role of e-h pair excitations. First, only non-reactive trajectories at a frozen surface are considered. While one trajectory is very close to the actual entrance channel for dissociation, this still can not capture possibly important non-adiabatic effects during the actual dissociation at a mobile substrate. Second, an adiabatic spin-transition from the triplet ground-state of the free O$_2$ molecule to the adsorbed singlet state is assumed. While a definite answer to the first point has yet to be found, there are good arguments that support the latter assumption. The high Pd density of states at the Fermi-level as well as the high Pd mass number favour both coupling mechanisms generally discussed to relax the spin selection rules that suppress reactions like oxygen dissociation, namely spin-orbit coupling and electron tunnelling between substrate and adsorbate. With respect to the latter it is noteworthy that an intriguing asymmetry of electron-hole pair excitations in the majority and minority spin channel is obtained in the present framework. The excess minority spin excitations span thereby an energy range around the Fermi-level in which there is a larger overlap of Pd substrate states with the minority $2\pi^{*\downarrow}$ orbital of the O$_2$ molecule. In the here implied adiabatic spin transition through efficient electron tunneling, this nicely illustrates the mechanism with which the non-spin polarized substrate electronic structure quenches the adsorbate spin-density.

\ack

Funding by the Deutsche Forschungsgemeinschaft in project RE 1509/7-1 is gratefully acknowledged. We thank Matthias Timmer and Peter Kratzer for valuable and stimulating discussions.

\section*{References}


\begin{thebibliography}{10}
\bibitem{meyer11_2}
Following Lorente and Persson \cite{lorente00b}, the use of ultrasoft pseudopotentials is properly considered in our \CASTEP\ implementation which evaluates \eref{eq:PDOS}.
\end{thebibliography}


\end{document}